\documentclass[sigconf]{acmart}
\acmYear{2021}
\acmConference{Preprint}{Under Review}{2021}
\renewcommand\footnotetextcopyrightpermission[1]{} 
\usepackage[utf8]{inputenc}
\usepackage[english]{babel}
\usepackage{natbib}
\usepackage{qtree}
\usepackage[ruled,vlined]{algorithm2e}
\usepackage{float}
\SetKwInput{KwInput}{Input}              
\SetKwInput{KwOutput}{Output}

\title{Dynamic Social Media Monitoring for Fast-Evolving Online Discussions}
\author{Maya Srikanth}
\affiliation{%
 \institution{California Institute of Technology}
 \city{Pasadena}
 \state{California}
}
\email{mayassrikanth@gmail.com}

\author{Anqi Liu}
\affiliation{
 \institution{California Institute of Technology}
 \city{Pasadena}
 \city{California}
}
\email{anqiliu@caltech.edu}

\author{Nicholas Adams-Cohen}
\affiliation{%
 \institution{Stanford University}
 \city{Palo Alto}
 \state{California}
}
\email{nadamsco@stanford.edu}

\author{Jian Cao}
\affiliation{%
 \institution{California Institute of Technology}
 \city{Pasadena}
 \state{California}
}
\email{jccit@caltech.edu}

\author{R. Michael Alvarez}
\affiliation{%
 \institution{California Institute of Technology}
 \city{Pasadena}
 \state{California}
}
\email{rma@caltech.edu}

\author{Anima Anandkumar}
\affiliation{%
 \institution{California Institute of Technology}
 \city{Pasadena}
 \state{California}
}
\email{anima@caltech.edu}

\begin{document}

\begin{abstract}
    Tracking and collecting fast-evolving online discussions provides vast data for studying social media usage and its role in people's public lives. However, collecting social media data using a static set of keywords fails to satisfy the growing need to monitor dynamic conversations and to study fast-changing topics. We propose a dynamic keyword search method to maximize the coverage of relevant information in fast-evolving online discussions. The method uses word embedding models to represent the semantic relations between keywords and predictive models to forecast the future time series. We also implement a visual user interface to aid in the decision making process in each round of keyword updates. This allows for both human-assisted tracking and fully-automated data collection. In simulations using historical \#MeToo data in 2017, our human-assisted tracking method outperforms the traditional static baseline method significantly, with 37.1\% higher F-1 score than traditional static monitors in tracking the top trending keywords. We conduct a contemporary case study to cover dynamic conversations about the recent Presidential Inauguration and to test the dynamic data collection system. Our case studies reflect the effectiveness of our process and also points to the potential challenges in future deployment.
\end{abstract}
\maketitle
\section{Introduction}

Conversations on social media platforms like Twitter are often dynamic \citep{bruns}. As new events occur, the language in tweets changes --- words rise and fall in popularity and evolve with time. Influencers, celebrities, and political leaders often change the topics discussed online to sell their products, brands, and ideas.  Furthermore, extremist groups, state-sponsored organizations targeting dissidents, and those intent on harassing and intimidating others online often permute the syntax of existing keywords or create new hashtags to avoid detection by social media platforms (and others who might be monitoring them) \citep{Badawy_etal_2018}. To understand the natural evolution of online discussions and detect abusive conversations in real-time, it is important to develop data collection methods which track shifts in online discourse.


Many researchers who use social media data from Twitter collect data using a set of static (unchanging) keywords and hashtags, e.g., \citep{cao2020reliable}.  But as previous research shows, static data collection methods fall short when social media conversations change, either because the language used to discuss some topic alters or the hashtags are syntactically  modified \citep{king_lam_roberts_2017, liu_etal_2019}.  Thus, there is a need for building dynamic social media monitors that can adapt to changes in social media conversations.

Developing a dynamic social media data collection monitor that can update keywords and hashtags is a challenging task.  Prior research has proposed methods that require human intervention, or are semi-automated \citep{king_lam_roberts_2017, liu_etal_2019}. Other researchers may prefer fully-automated methods.  Either way, a dynamic monitor requires the integration of a number of different methods:  it needs to start with a collection of social media posts on a certain topic, which can then be analyzed by natural-language processing tools to determine if there are new keywords or hashtags emerging in the data over time.  The dynamic monitor then needs a predictive modeling step, where it forecasts the likelihood that the new language on the topic will continue to grow.  Finally, based on the predictive model, the dynamic monitor then needs to adjust the keywords or hashtags it collects information on, and needs to continue to analyze whether new keywords or hashtags should continue to be included in the monitor.  
\begin{figure*}[t]
\centering
\begin{tabular}{c}
       
        \includegraphics[width = 0.7\textwidth]{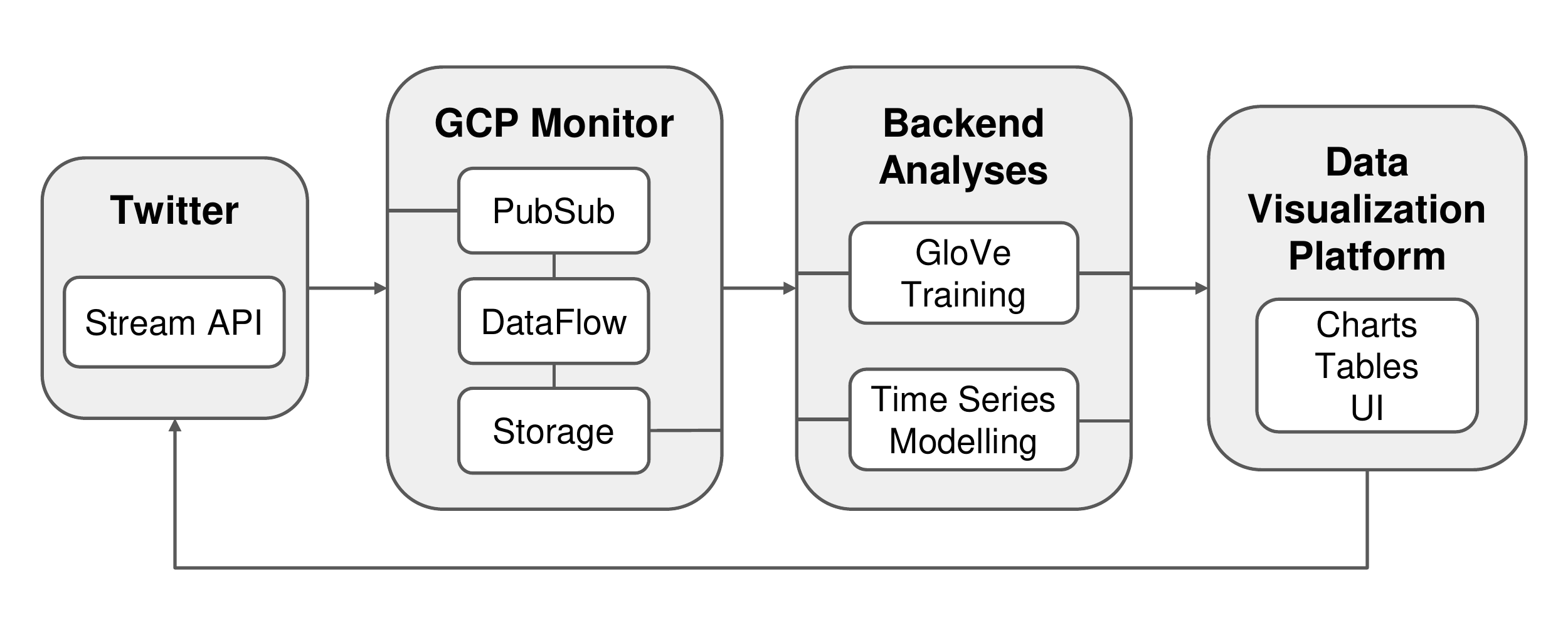}   \\
         
\end{tabular}
\caption{Workflow of our Data Collection, Storage, Analysis, and Visualization Platform. Here we use Twitter APIs as an example for data collection and we use Google Cloud Platform (GCP) as an example for data streaming and storage. GCP Monitor works in a sequential way while the backend analyses can be parallel.}
\label{fig:flowchart}
\end{figure*}

We design and implement a dynamic monitor for collecting data on fast-evolving online discussions. We allow for both semi-automatic and fully-automatic data collection. Our final dynamic monitor design uses word embeddings, corpus frequencies, and predictive time series modelling to visualize trends in a real-time social media discussion, recommend new keywords for data streaming, and facilitate social media data collection.  We provide the code so that other researchers can use these tools \footnote{https://github.com/mayasrikanth/DynamicMonitor}.  Figure \ref{fig:flowchart} demonstrates the four components  of our framework, which include data collection and storage, data analysis, and visualization.

\begin{figure*}[htbp]
\centering
\begin{tabular}{c}
        \includegraphics[width = 0.9\textwidth]{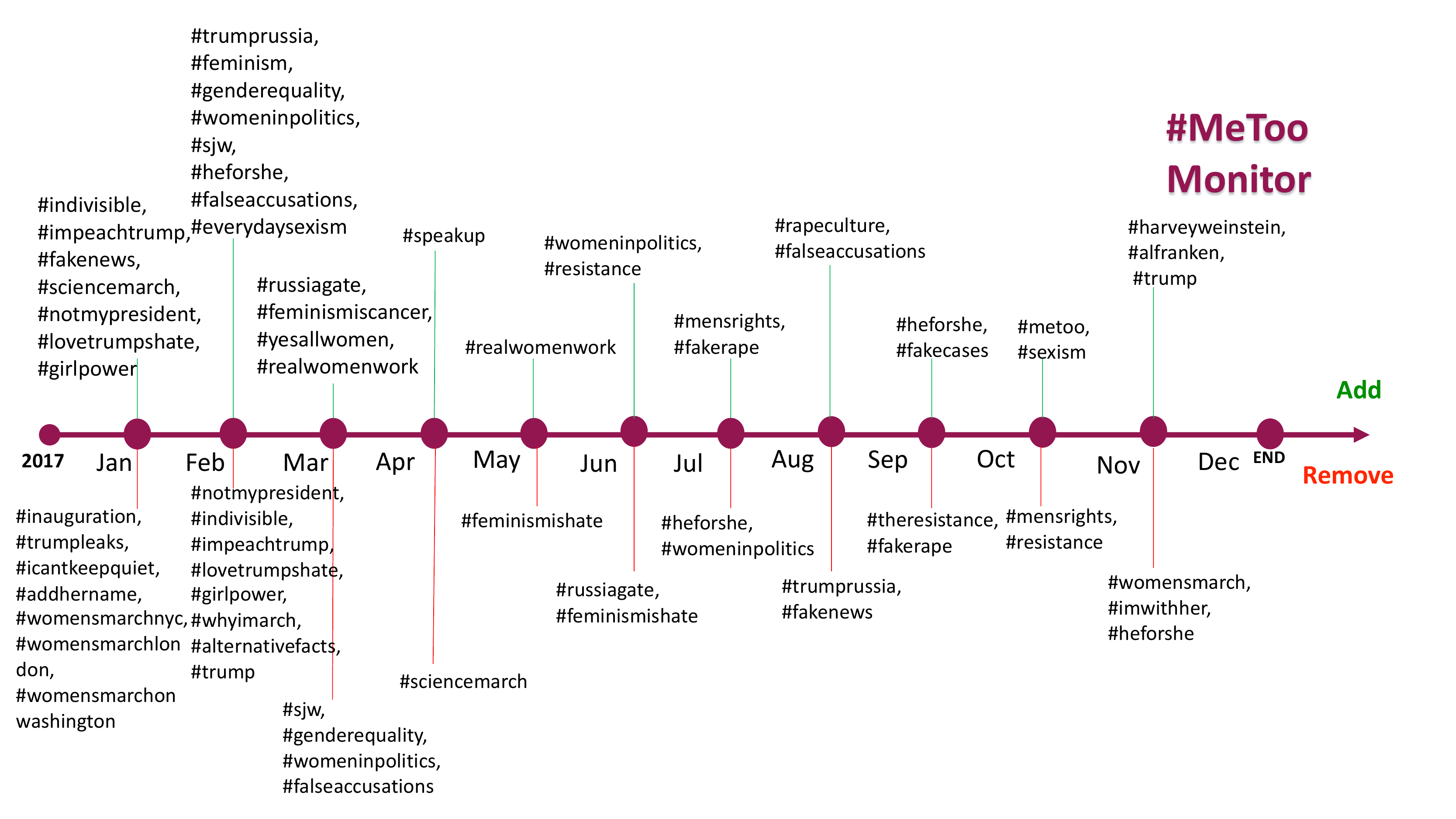} 
\end{tabular}
\caption{Keywords update (add and remove) by the dynamic monitor for data collection in historical \#MeToo simulation. The evolving keyword set aligns with real-world events: throughout early to mid 2017, the monitor tracks important movements (e.g. `\#womensmarchoonwashington in Jan 2017 and `\#sciencemarch' in Apr 2017) and gender-equality hashtags (`\#womenwhowork', `\#womeninpolitics). In late 2017, the monitor picks up the viral \#MeToo movement with `\#metoo' in Oct, as well as related figures like `\#harveyweinstein' and `\#alfranken'. The monitor also picks up traditionally anti-\#MeToo hashtags throughout the simulation, with `\#sjw', `\#feminismishate', `\#mensrights', `\#fakecases'. }
\label{fig:MeTooKeywords}
\end{figure*}

\begin{figure*}[htbp]
\centering
\begin{tabular}{c}
        \includegraphics[width = 0.9\textwidth, trim={0cm, 4cm, 0cm, 0cm}, clip]{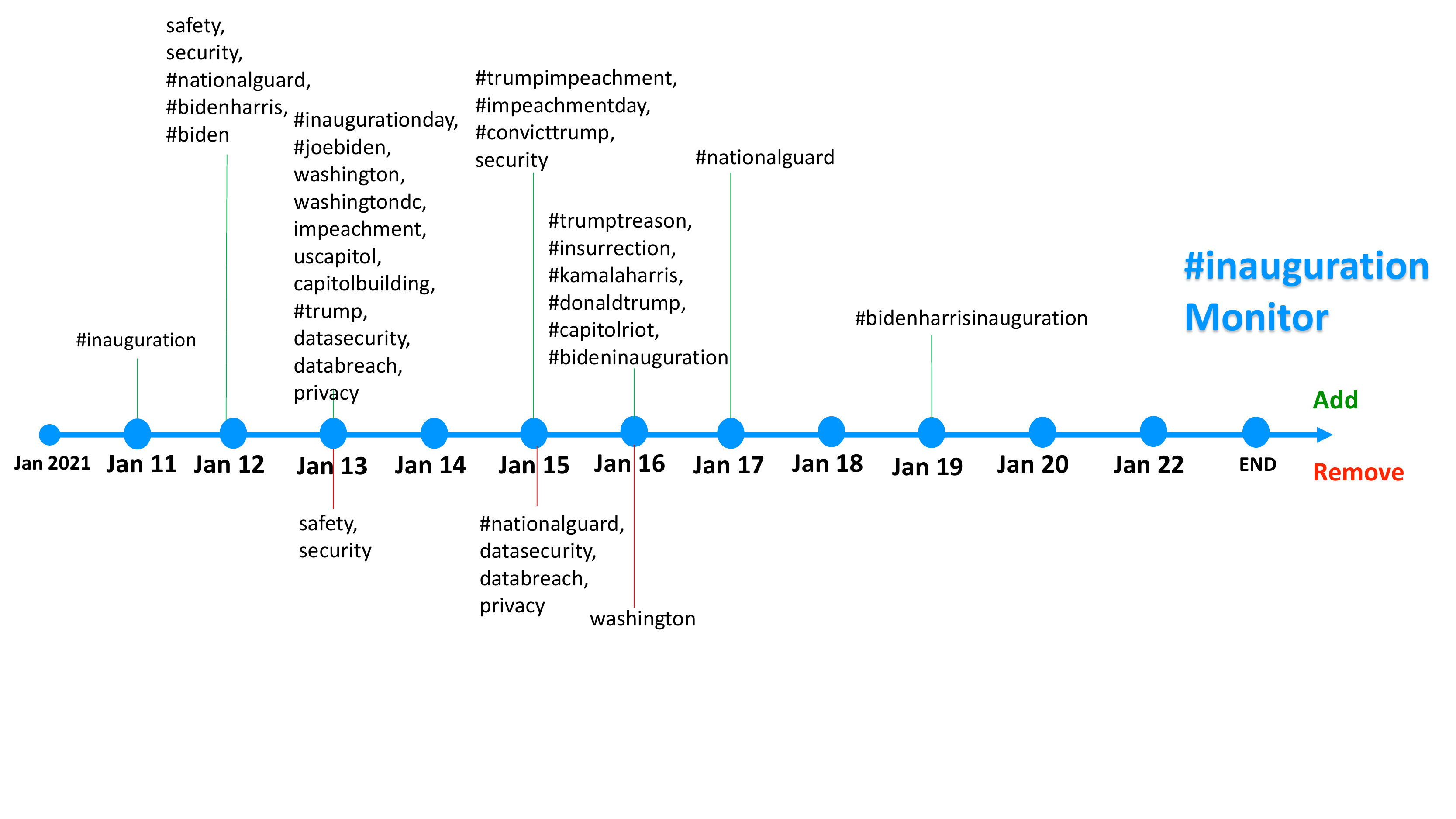}   
          
\end{tabular}
\caption{Keywords update (add and remove) by the dynamic monitor for data collection in real-time case study of 2021 \#inauguration discussions on Twitter. Emergence of `\#nationalguard' on Jan 12 likely corresponds with the 2021 capitol protests (as does `\#capitolriot' and `\#capitolbuilding' on Jan 16), and emergence of `databreach' on Jan 13 is likely linked to the hacking of Parler's data. As time goes on, we see that the monitor begins to focus on politicians, detecting various forms of `\#biden' (`\#joebiden', `\#bidenharris' `\#bideninauguration', `\#bidenharrisinauguration'), `\#trump' (`\#donaldtrump'), and `\#kamalaharris'.  After inauguration on Jan 20th, all \#inauguration discussions naturally lost traction (see Figure \ref{fig:casestudy1} for declining frequencies after the fact).}
\label{fig:InaugurationKeywords}
\end{figure*}

Our work makes the following contributions: 
\begin{enumerate}
    \item By combining word embeddings with predictive time series modeling, our methodology allows for fully-automated, semi-automated, or completely by-hand updating of keywords used to pull social media data over extended periods. 
    \item By providing an online interface, we give analysts the means to visualize the various components of the dynamic monitor: using our code, researchers can oversee the operations of a dynamic monitor and alter the course of their data collection.
    \item By conducting simulations and case studies with both  historical data from the prominent 2017 \#MeToo movement and the tumultuous 2021 Presidential Inauguration, we put the deployed components of our process into use on real-world, fast-evolving online discussions to demonstrate their efficacy.
\end{enumerate}

More specifically, Figure \ref{fig:MeTooKeywords} summarizes our dynamic monitor updates of keywords in the simulation, where we only use GloVe embeddings and keyword frequency information to model the keywords relations and apply human judgement for keywords updates. Our method achieves 37.1\% higher F-1 score on average than the traditional static monitor in tracking the top trending keywords for each month in 2017 (Table \ref{fig:F1Table}). Figure \ref{fig:InaugurationKeywords} shows dynamic monitor updates of keywords used to collect real-time \#inauguration data in 2021.


\section{Task Formulation}

\paragraph{Notation} We define $s_t$ as the set of keywords we are interested in tracking at timestep $t$. Using $s_t$, we can filter a corpus $K_t$ using APIs from different social media platforms. For example, Twitter provides various APIs that filter tweets containing specific  keywords and hashtags. We define $G_t$ as the semantic representation of words included in the filtered corpus. In prevalent word embedding models, we can use a vector to represent each word's semantic relation with other words. We use $P_t$ to represent the future trends for each word. $P_t$ can be directions (increases or decreases) or specific frequencies. Our goal then is to build a system for dynamic data collection such that maximizes the information coverage for evolving discussions around certain topics. As such, given keywords $s_t$ and the corresponding corpus $K_t$, we aim to update the keyword set $s_{t+1}$ according to the patterns in $G_t$ and $P_t$. 

\paragraph{Data Collection and Storage} We assume data collection using the APIs from social media platforms can be conducted efficiently. In practice, there exist additional difficulties in conducting this data filtering in a large scale and storing the data reliably \cite{cao2020reliable}. However, in this paper, we focus on the process of decision making for updating the keywords and the visualization for facilitating the decision making when human intervention is needed. 

\paragraph{Decision Making on Updating Keywords} Based on $K_0,...,K_t$, we first need to generate the semantic representation $G_t$ and the future trend $P_t$.  Given $G_t$ and $P_t$, there are different many ways to determine the updated keywords, and in many contexts it is desirable that this updating process remain flexible. When there is a smooth trend in topic shifting or there are emerging events that slowly change the direction or sentiment of the discussion, fully-automatic updating using simple rules is sufficient. However, in some contexts human intervention guided by information of $G_t$ and $P_t$ is needed. For example, swift topic shifting is hard to forecast using historical data. In these situations, human intervention may be helpful. 

\paragraph{User Interface} Demonstrating $G_t$ and $P_t$ for decision making is important for closing the loop for future data collection. The interface should consist of user-friendly elements that enable clear illustration of the semantic relation between keywords and the forecasts of future trends.

\section{Method}

At present, we have built and deployed the word embedding component of our dynamic keyword and hashtag monitoring process (focusing at this point on hashtags). The predictive modeling component of our process is still under development, as we discuss later, we have implemented some time series predictive models and will demonstrate their utility in future research.

\subsection{Word Embeddings to Model Keyword Relations}

At time $t$, we train the GloVe model \cite{pennington2014glove} using $K_t$ to produce 50-dimensional word embedding representations $G_t$ of all tokens in our filtered corpus $K_t$. 
GloVe can represent linear substructures in data. It is a log-bilinear model with a weighted least-squares objective, and aims to learn word vectors such that their dot product equals the logarithm of the words’ probability of co-occurrence. In the resulting word vector space, cosine similarity indicates linguistic or semantic similarity between two words, while vector differences capture analogies between pairs of words. These embeddings allow us to project each token into euclidean space, where we can use distance metrics to measure "closest" neighbors to each of our keywords $s \in s_t$.

\subsection{Predictive Method to Forecast Keyword Frequency Trend }

Our latest implementation supports time series forecasting with ARIMA (auto-regressive integrated moving average) for univariate frequency data. Specifically, we visualize keyword frequencies and predict their trajectory within a confidence interval to determine whether a discussion topic is increasing or decreasing. 

Prior to any forecasting, we apply log transform to all corpus counts in order to stabilize the variance \citep{Lutkepohl2012} and induce stationarity in all series by differencing lags. For each keyword, we grid search ARIMA model parameters and select those which maximize performance (minimize mean-squared error) on the validation set. The best model is then used to forecast frequencies $10-15$ time-steps into the future. 

While linear models like ARIMA do not outperform larger deep learning models when data is abundant, they can be informative in earlier stages of data streaming when textual data is sparse. For applications with more abundant data, we plan to include options for training more robust deep learning models in the future. 

In order to collect enough data to allow for meaningful time series prediction during the data collection process, our dynamic implementation design pulls $7$ days worth of data from Twitter's REST API using the starting set of keywords as the query.

\subsection{Proposed Algorithm}
With $G_t$ and $P_t$ defined above, our algorithm proceeds as follows.

For each $s \in s_t$ at time $t$, from embedding space $G_t$ we find $C_{30}$, the set of 30 closest neighbors to keyword $s$, defining closest with the cosine similarity metric. From this set of thirty words, we choose the most relevant hashtags or mentions in the domain or event we are tracking. We define these neighbors as $n_s \in C_s \subseteq C_{30}$.

For each $n_s \in C_s$, we use our time series model $P_t$ to predict future frequencies for the hashtag or mention. If we predict $n_s$ is declining in future time periods, we drop these values from the set $C_s$. We define $C_{s'} \subseteq C_s$ as the set of neighbors \emph{without} declining time series predictions.

Finally, define $s_{t+1}$, the set of  keywords we track in the next time period, as $C_{s'}$. 

 \begin{algorithm}[hbt!]
\DontPrintSemicolon
    \KwInput{$s_t$: keyword set at $t$, $K_{t}$: filtered corpus at $t$}
    \KwOutput{\newline $s_{t+1} = \{\}$}
    \KwData{\newline $G_{t} \leftarrow$: obtain 50-dimension GloVe embeddings trained on $K_{t}$. $P_{t} \leftarrow$: update time series models with latest frequency data from corpus} 

    \For{$s \in s_{t}$} {
        {1. $C_{30}$: 30 closest neighbors to $s$ in embedding space $G_{t}$.}\\
        {2. $C_{s}$: choose relevant hashtags or mentions from $C_{30}$ } \\
        {3. $C_{s'}$: Discard hashtags from $C_{s}$ with declining trend lines in time series prediction or low corpus counts.}\\
        {4. $s_{t+1} \longleftarrow C_{s'}$}
    } 
    \textbf{Return} ${s_{t+1}}$. 
  \caption{Dynamic Algorithm}
  \label{humaninput}
 \end{algorithm}

 \subsection{Platform Features vs. Case Study Methods}
The case studies we conduct on historical \#MeToo data and real-time 2021 \#inauguration discussions use a simplified version of the algorithm above. In both studies, the human-assisted decision process to drop or add keywords is informed \textbf{only by corpus frequencies and GloVe embeddings}: we do not use time series models to 
predict the trajectory of keywords. 

However, we observed from independent analysis that even linear time series models can produce reasonable short-term estimates of keyword frequency trajectories. Thus, our data visualization platform code includes interactive charts with time series forecasts (see Figure \ref{fig:interface2}), as well as code for training ARIMA on real-time Twitter data. 

\section{Implementation}
\subsection{Data Visualization Platform}

To enable researchers to collaboratively adapt their social media data collection process to dynamically changing discussions, we are building a data visualization platform which integrates a browser-based user interface on the frontend with scripts for data streaming and AI-driven predictive modelling on the backend.  Below, we detail the features and capabilities of this tool, before demonstrating its use on real-time monitoring of 2021 Twitter discussions about \#inauguration.

\begin{figure}[H]
\begin{tabular}{cc}
           \includegraphics[width = 0.47\textwidth, trim={5cm, 1cm, 0.2cm, 5cm}, clip]{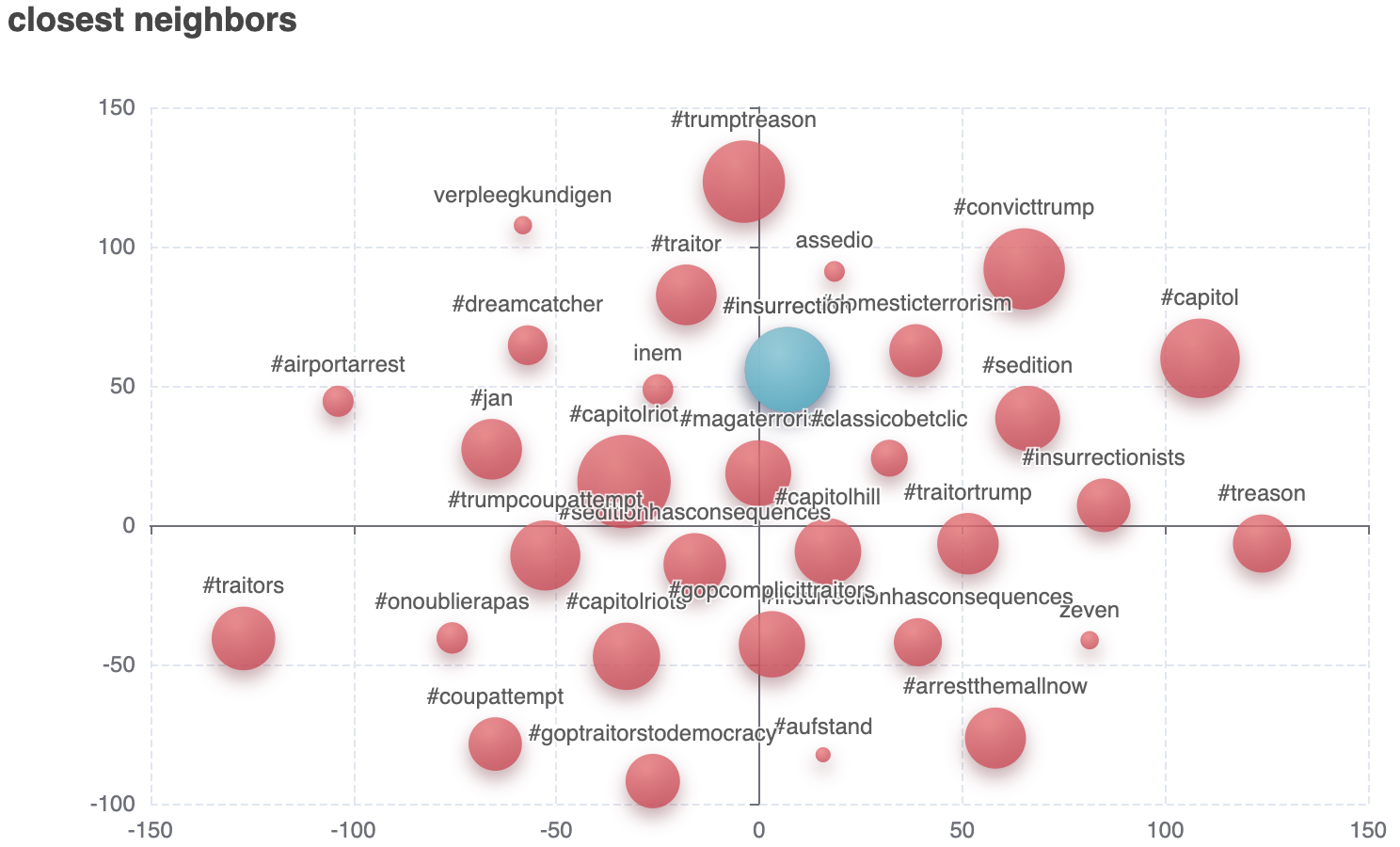}   
 
\end{tabular}
\caption{Interactive tsne (t-distributed stochastic neighbor embedding) plot of the closest 30 neighbors to the tracked keyword \#insurrection (shown in blue). Size of bubble corresponds to corpus frequency of keyword. }
\label{fig:interface1}
\end{figure}

\begin{figure}[H]
\begin{tabular}{cc}
           \includegraphics[width = 0.47\textwidth, trim={0.3cm, 0cm, 0.2cm, 0.3cm}, clip]{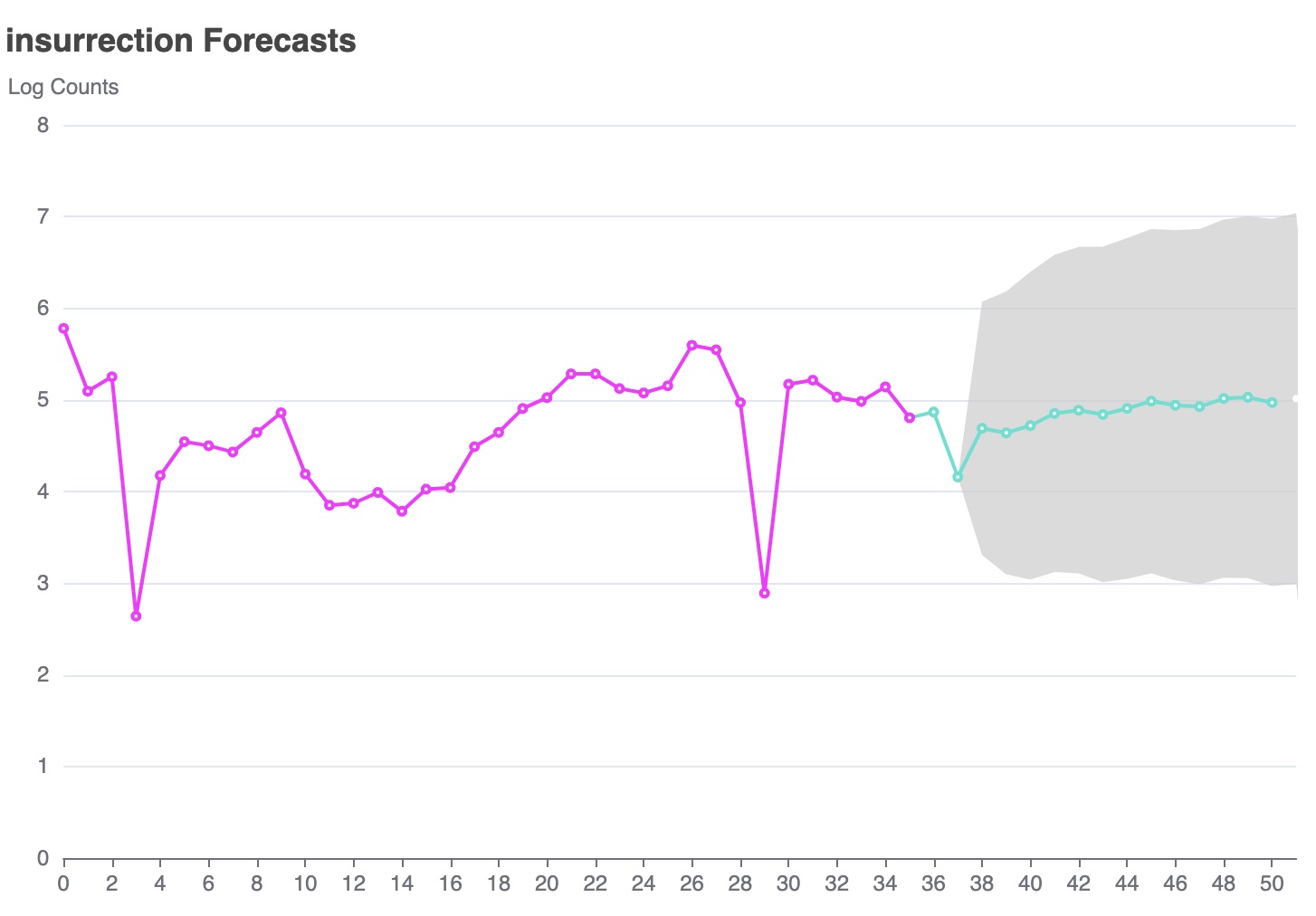}   
 
\end{tabular}
\caption{Forecast plot generated after training ARIMA on log-transformed \#insurrection frequency data, which predicts log frequencies 15 timesteps into future and provides a 95\% confidence interval. ARIMA $p,d,q$ values were naively grid-searched. MSE for forecasts is $0.259$. X-axis shows last 50 hours in case study. }
\label{fig:interface2}
\end{figure}

\begin{figure}[hbt!]
\begin{tabular}{cc}
         \includegraphics[height=5.0cm, scale = 0.60, trim={0.5cm, 0.1cm, 1.5cm, 0.5cm}]{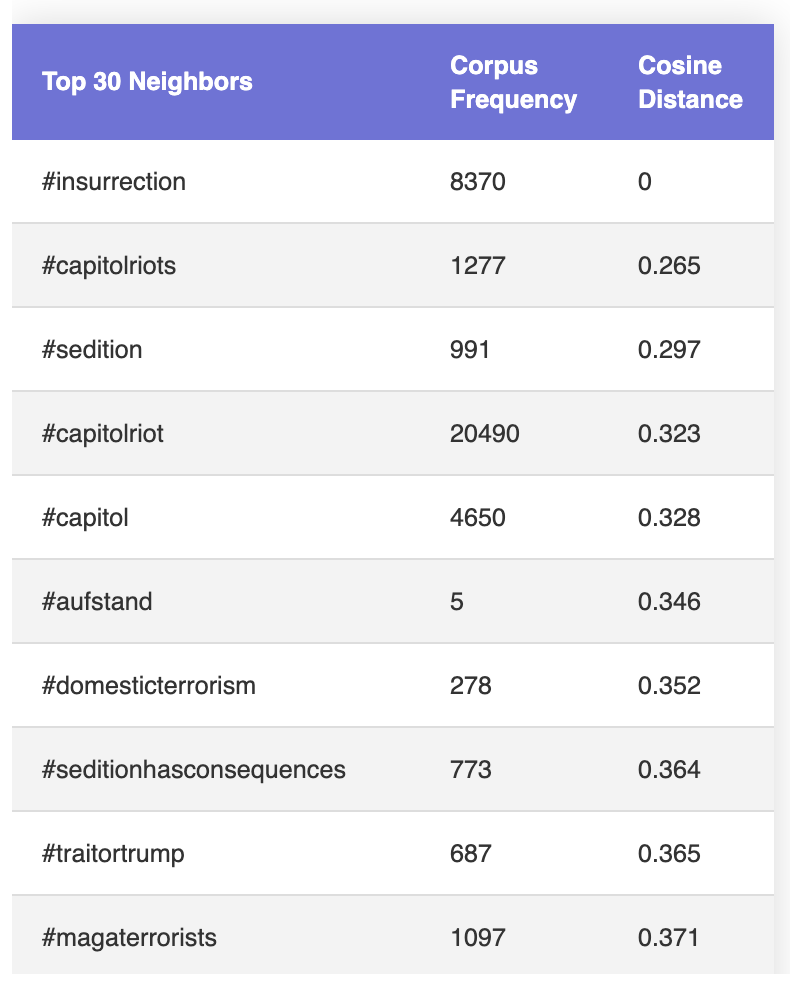} 

\end{tabular}
\caption{Table of 30 closest neighbors to \#insurrection, sorted by a linear combination of keyword cosine distance to \#insurrection and corpus frequency.}
\label{fig:interface3}
\end{figure}

\paragraph{Frontend}
The frontend of our platform is built with JavaScript, HTML, and CSS. We use the JavaScript library echarts to render figures and tables which are dynamically updated by the backend to reflect real-time data. Our frontend code uses the echarts API to make our figures interactive: users can ``zoom in" on specific numerical values. We host our frontend using Github pages and plan to release our code to allow any research group to host their own personalized version of the data visualization platform. Figures \ref{fig:interface1}, \ref{fig:interface2}, and \ref{fig:interface3} show the interface visualizations for \#insurrection, a tracked keyword in our 2021 \#inauguration case study. We provide a demo of our frontend UI: https://mayasrikanth.github.io/social-media-trends/index.html.

\paragraph{Backend} 

Our code streams data using Twitter API and stores it in a cloud compute service (Oracle Cloud or GCP). On the cloud, several scripts preprocess data, train GloVe and other predictive time series models, and update the frontend interface. Then on the frontend, the user can select new keywords or drop old ones to customize their data collection process. Time intervals for updating all frontend visualizations are customizable, with a lower bound of $\approx 15$ minutes.

\subsection{Predictive Model Guided Decision Making}

Our dynamic monitor design lends itself to semi-automated and fully -automated data collection processes.

\paragraph{Semi-automated}
We consider a semi-automated data collection process to mean a human-assisted one. That is, a group of researchers interested in tracking a particular set of topics on social media can utilize the AI-driven keyword recommendations on the frontend to alter their keyword set throughout data collection. The time series forecasting and word embeddings can uncover new discussion topics and indicate whether existing hashtags are increasing or decreasing in frequency: given this real-time information, researchers can  adjust their data streaming. 

\paragraph{Fully-automated}
Our implementation leaves room for a fully-automated approach which sorts candidate keywords using a linear combination of predictive factors, such as:
\begin{equation}
    s_{i} = \alpha \cdot m_{i} + \beta \cdot d_{i} + \gamma \cdot f_{i} + \delta \cdot v_{i}
\end{equation}

where $s_{i}$ is the ``virality" score for a given keyword, $m_{i}$ is the slope of the projected frequency trend-line, $d_{i}$ is the average cosine distance from the current set of keywords, $f_{i}$ is the current corpus frequency of the keyword, and $v_{i}$ is the variance of the keyword's frequency trend-line. Keywords can be sorted according to this metric, and the first $3-5$ keywords can be automatically added to the set. Further, removal criterion can be imposed--e.g. we can drop keywords that are relatively old and have low usage in the corpus. The scaling factors for these variables can be customized to fit the research objective: for instance, if a researcher wants to track niche topics with low predicted popularity, they can reduce the weight of $f_{i}$. 

We take the semi-automated approach in our studies, as it better fits our research objectives of testing a human-in-the-loop dynamic data collection method. In future iterations, we will include framework for the fully-automated keyword selection in our code. 


\section{Evaluation}
\begin{figure}[hbt]
\begin{tabular}{l}
           \includegraphics[height=5cm, scale = 0.30, trim={2cm, 0.09cm, 0.9cm, 0.5cm}]{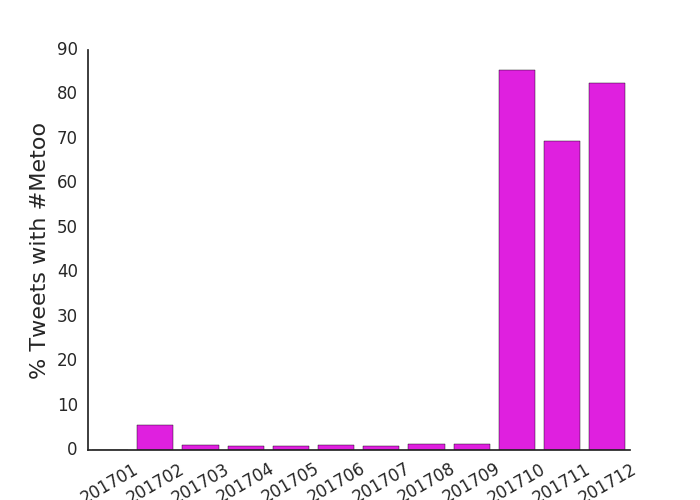} \\
           \\
           \includegraphics[height=5cm, scale = 0.30, trim={2cm, 0.09cm, 0.9cm, 0.5cm}]{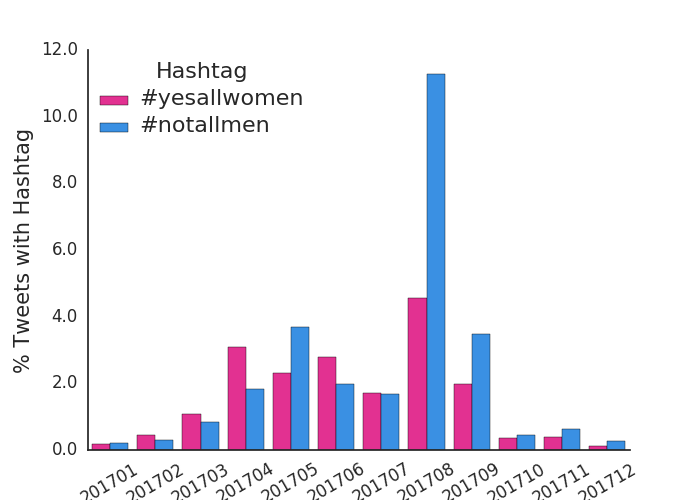} \\
           \\
\end{tabular}
\caption{{\bf Upper}: {\bf Upper}: Percentage of monthly tweets containing \#MeToo, with an unexpected surge in October 2017. {\bf Bottom}: Percentage of tweets with "\#yesallwomen" (pink) or "\#notallmen"(blue) throughout 2017.}
\label{fig:totalcounts}
\end{figure}
\subsection{Simulation Designed using \#Metoo Data}

\paragraph{\#MeToo Data:}  Building on a series of women's rights marches and protests, the  \#MeToo movement went viral on Twitter in October 2017 after media outlets widely publicized sexual assault allegations against Harvey Weinstein. Women and men across the globe adopted the \#MeToo hashtag to stand in solidarity against sexual harassment and bring to light gender inequality issues across many spheres, from Hollywood to the tech industry. To study this evolving set of issues, our research team obtained a large collection of Twitter data about the \#MeToo movement;  this data was obtained directly from Twitter through Boolean filtering and is publicly available information.  Demonstrating the scale of women's rights discussions and the shift to \#MeToo, we show the total number of tweets in our data, as well as the percentage of these tweets that contain ``\#MeToo'' in Figure \ref{fig:totalcounts}.

\begin{figure}[hbt!]
\setlength{\tabcolsep}{0pt}
\centering
\begin{tabular}{cc}
\includegraphics[scale=0.45, trim={1.2cm, 1cm, 1cm, 1cm}, clip]{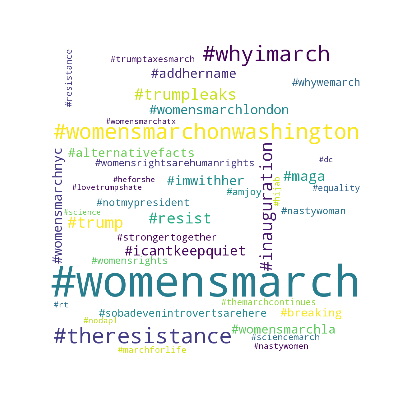}&   
\includegraphics[scale=0.45, trim={1.2cm, 1cm, 1cm, 1cm}, clip]{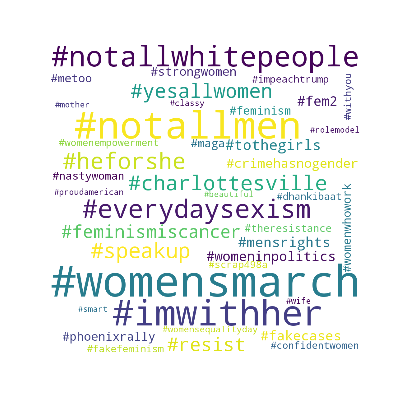} \\
(a) & (b)
\end{tabular}
\caption{(a) Word cloud containing 40 most frequently used hashtags in January 2017; (b)
Word cloud of 40 most frequently used hashtags in August 2017. The size of the words indicate the frequency. There is a significant topic shift on Twitter between these two months.}
\label{fig:janclouds}
\end{figure}

\paragraph{Evolution of Topics:} As shown in Figure \ref{fig:janclouds}(a), January 2017 sees an upsurge of the hashtag ``\#womensmarch'', which can be attributed to the Women's March on Washington, the U.S.'s largest single-day public demonstration that took place the day after President Trump's inauguration on the 20th to promote gender equality and civil rights. The controversial 2016 U.S. presidential campaign inspired many partisan discussions in January, with politically contentious hashtags like ``\#theresistance'', ``\#notmypresident'', ``\#trumpleaks'', ``\#alternativefacts,'' and ``\#fakenews''. The months February through September witness similar lines of discussion about politics, gender inequality, and feminism, albeit with evolving top terms. For instance, come June 2017, hashtags like ``\#imwithher'', ``\#everydaysexism'', and ``\#feminismiscancer'' rise to the forefront of online conversation.  As this discourse gains momentum, we see the emergence of movements and their corresponding anti-movements: take for instance ``\#yesallwomen'' (expressing prevalence of sexual harassment faced by women) vs. ``\#notallmen'' (a hashtag to defend males), both which are prominent in August 2017, as shown in Figure \ref{fig:janclouds} (b). Come October 2017, high-profile sexual assault allegations including those against Harvey Weinstein catalyzes the ``\#MeToo'' movement and spurs others to come forward with allegations against Larry Nassar, Kevin Spacey, Roy Price, and more over the next few months.

\paragraph{Simulation of a Dynamic Monitor:}  The shifts in the language used to describe a general set of issues and the advent of \#MeToo represent an ideal scenario to utilize a dynamic keyword algorithm. While theoretically we could have run our dynamic keyword algorithm in real-time, in order to better test our method, we simulate it on historical data from January 2017 to December 2017 to evaluate how well our algorithm tracks the evolving \#MeToo movement.  The aim of this simulation is to capture the most frequent keywords for each month starting in January 2017 and ending in December 2017. We simulate our dynamic keyword search process as follows: starting in January, we filter the overall corpus of historical January data with our seed keywords. Note that by filtering down the large set of historical data, we simulate gathering data from the Twitter API. We then analyze this filtered set with our dynamic keyword approach to determine a new round of keywords for the subsequent month. In this simulation, the top hashtags each month in the full \textbf{unfiltered} historical data represents the ground truth target set for evaluation.

\subsection{Evaluation Baselines and Metrics}
We compare our method with two baselines: Static and Last-Top.
\begin{enumerate}
    \item Static: uses the top $n$ keywords from January 2017 for each of the following months throughout the simulation.
    \item Last-Top: assumes previously trending hashtags can expose trending conversations in the current month and uses the keywords set from the last month.
\end{enumerate}

To make comparison between the methods more tractable, we set $n=15$ keywords at any time $t$, although in practice it is certainly possible to use more.

Since we have access to the top trending hashtags for each month in the whole dataset, we set these to be the ground truth, and evaluate the most frequently used hashtags pulled by various monitors against this. We use the Jaccard similarity index and F1-score for the evaluation. 

To calculate F1-score, we regard the proportion of correctly retrieved top trending hashtags in the retrieved hashtags ($R$) and in the ground truth hashtags ($G$) as precision and recall.

For the Jaccard similarity, we look at the union and the intersection of the retrieved hashtags ($R$) and the ground truth hashtags ($G$). Importantly, all monitors begin with the same set of keywords--the top $15$ most frequently used hashtags in our January 2017 \#MeToo data. 
\begin{align}
    F_1 = 2 \cdot \frac{\text{Precision}\cdot \text{recall}}{\text{Precision}+ \text{recall}}
\end{align}

\begin{align}
    J = \frac{R \cap G}{R \cup G}
\end{align}


\subsection{Result Analysis}
\begin{figure}[h]
\centering
\begin{tabular}{cc}
           \includegraphics[height=5.5cm, scale = 0.40, trim={1cm, 0.1cm, 1.5cm, 0.5cm}]{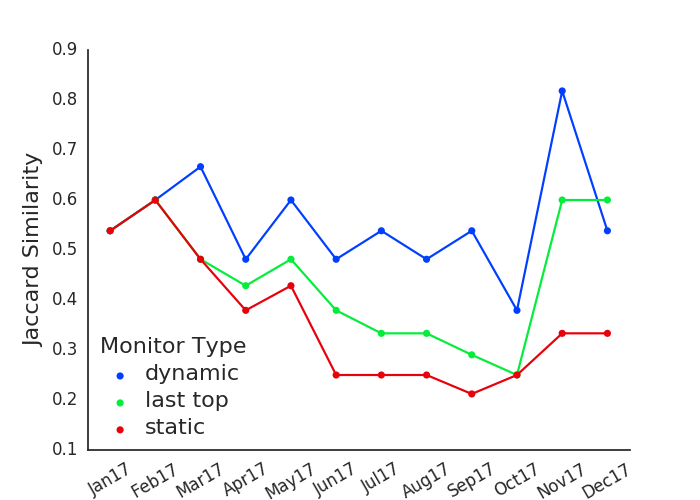}   \\
           \\
           \includegraphics[height=5.5cm, scale = 0.40, trim={1cm, 0.1cm, 1.5cm, 0.5cm}]{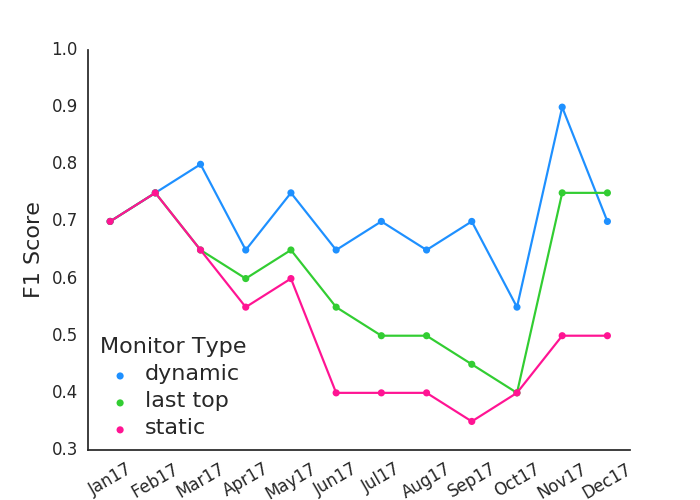}
\end{tabular}

\caption{The similarity between the set of top 20 hashtags in the subset of data recovered by each algorithm (dynamic, last-top, and static) and the target set of top 20 hashtags in the full month of data in terms of Jaccard index (top) and F1 score (bottom). Dynamic algorithm outperforms the baselines at most timesteps. In particular, the dynamic monitor covers 75\% more top trending keywords than the static monitor on average.} 
\label{fig:results}
\end{figure}

 \begin{table}[hbt!]
 \caption{Global Performance Comparison}
 \label{tab:global}
 \centering
 \begin{tabular}{|l|l|l|l|}
\hline
 & {\small  Jaccard } & {\small Avg. F1} & {\small Avg. F1} \\
 & & {\small Weighted} & {\small Unweighted} \\ 
 \hline
Dynamic & $ .5406$  & $.6976$ & $.7083$  \\
Last-Top  & $.508$ & $.6665$ & $.6041$ \\
Static & $.4594$ & $.6199$ & $.5166$\\
\hline
 \end{tabular}
 \label{fig:F1Table}
\end{table}

Figure \ref{fig:results} shows the performance of our algorithms and the baselines. Table \ref{tab:global} shows the weighted average of Jaccard similarity index and F1-score of each algorithm as percentages with respect to the to target set (top hashtags for the full-month data), where the weight is the proportion of the size of monthly data to the entire corpus size. We also compare with an unweighted average F1-score. In all of the metrics, our method outperforms the Last-Top baseline and the conventional Static monitor typically used in social science research. This indicates that our method can better capture the trending topics in the \#MeToo data without losing track of the newly emerging keywords and hashtags. \newline

For a more granular analysis, the fact that all monitors start with the same keywords explains identical performance in the first month. The Dynamic and Last-Top monitors perform most poorly in October 2017: this is due to the fact that neither picks up the sudden emergence of \#MeToo. Through preliminary analysis, we find that this dramatic surge in usage of \#MeToo occurs within a few hours, and that even powerful deep learning models perform poorly in prediction tasks with this dataset-specific anomaly.

\paragraph{Keyword Evolution in \#MeToo Simulation:} Figure \ref{fig:MeTooKeywords} demonstrates the keywords retrieved by our algorithm, which reflects the evolution of the topics in the early stage of the \#MeToo movement.  

\section{Case Study:  President Biden's 2021 Inauguration}

To provide a contemporary case study using the dynamic keyword selection method, we started the dynamic monitor on January 11, 2021 (at 10:40:44 Pacific Standard Time), with a single keyword, ``inauguration.''  We used the Twitter data collection architecture developed by \cite{cao2020reliable} to stream data. We stopped collecting data using this dynamic monitor on January 22, 2021 at 15:21:06 Pacific Standard Time.  

\begin{figure}[htb!]
\centering
\begin{tabular}{cc}
       
           \includegraphics[width = 0.45\textwidth, trim={0.65cm, 0.7cm, 0.2cm, 0.2cm}, clip]{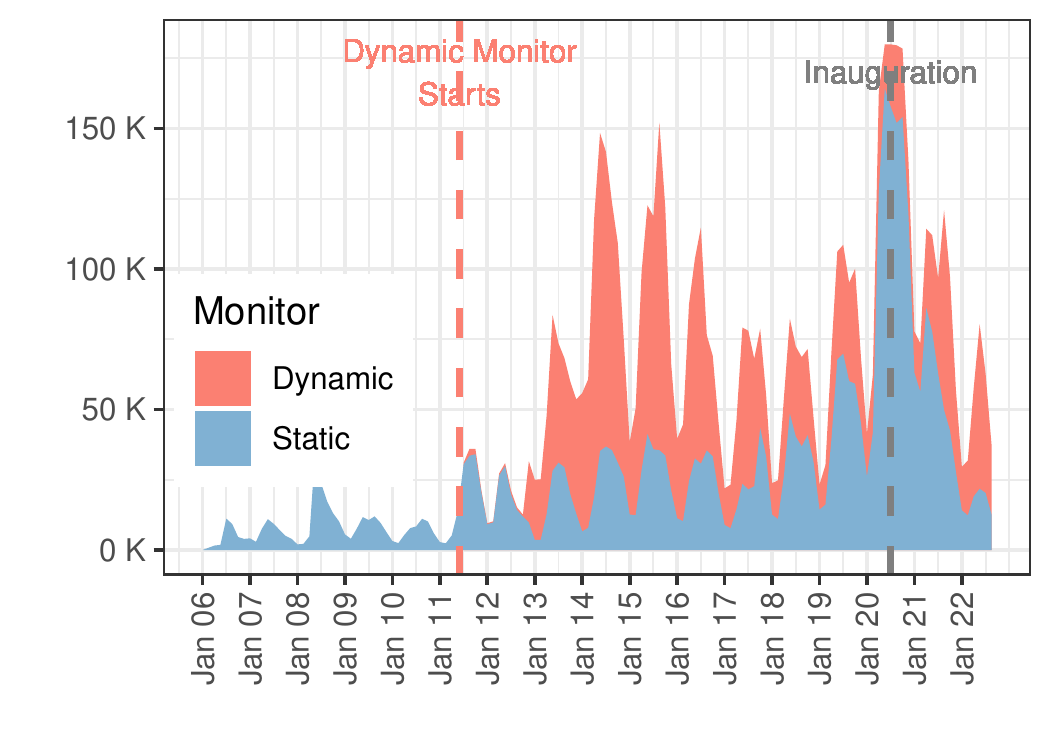}   \\
\end{tabular}
\caption{Hourly Number of Tweets containing \#inauguration from Jan 06, 2021 to Jan 22, 2021.}
\label{fig:casestudy1}
\end{figure}

We show in Figure~\ref{fig:casestudy1} the amount of data collected by this dynamic monitor (by hour in the top panel, and by day in the bottom panel).  The bottom panel of Figure~\ref{fig:casestudy1} provides what a static monitor would have collected, daily, during this period. During the first full day of data collection, the dynamic monitor pulled 499,111 tweets; rising by January 14, 2021 to 2,470,410 on that day.  The number of tweets collected by the dynamic monitor peaked on January 20, 2021 (the day of the inauguration), pulling 3,434,650 tweets.  In total, the dynamic monitor collected 19,723,508 tweets.    

\subsection{Data Collection and Update Procedure}

We pulled discussions related to \#inauguration from Twitter's historical and streaming APIs. In order to adapt the keyword set to account for dynamically changing conversations, we utilized our data visualization platform in a human-assisted update procedure. Specifically, on each day of the experiment at $~5pm$ PST, we used the closet neighbor table and tsne plots on our webpage (see \ref{fig:interface}, \ref{fig:interface1}) to collaboratively determine whether to add or subtract keywords for data collection.

\subsection{Comparison with the Static Monitor}
\begin{figure}[hbt!]
\setlength{\tabcolsep}{0pt}
\centering
\begin{tabular}{cc}
           \includegraphics[height=5cm, scale = 0.40, trim={2cm, 0.1cm, 1.5cm, 0.5cm}]{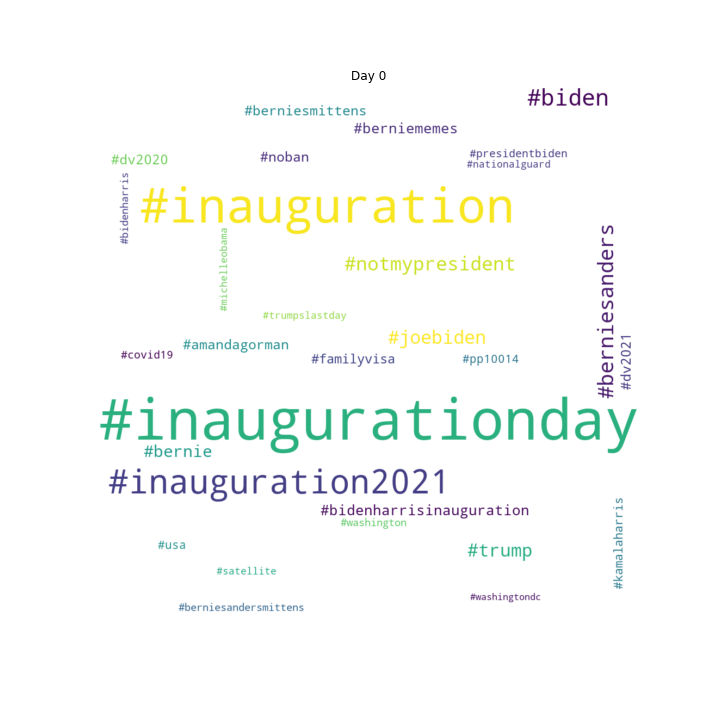}   &
           \includegraphics[height=5cm, scale = 0.40, trim={2cm, 0.1cm, 1.5cm, 0.5cm}]{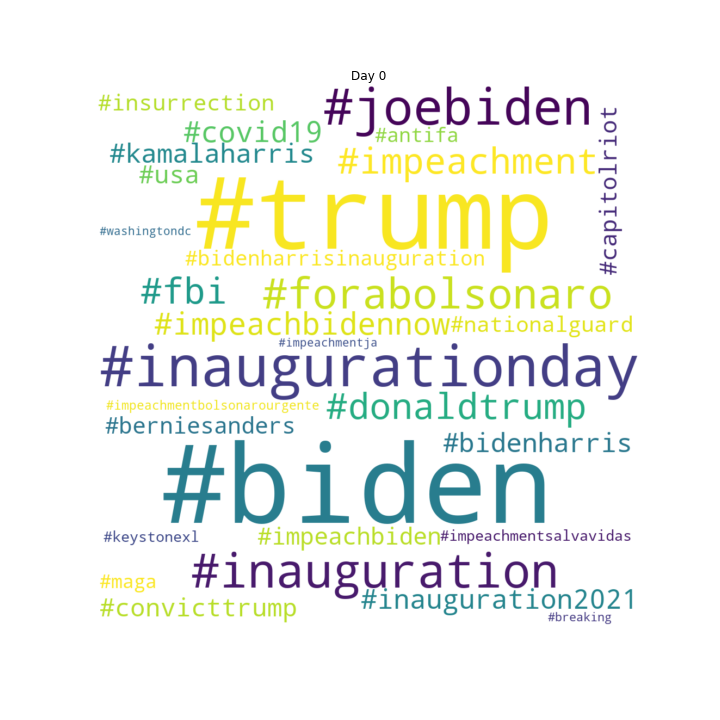} 
\end{tabular}
\vspace{-20pt}
\caption{Word cloud from the static monitor (left) and the dynamic monitor (right) on day 9 of the experiment showing the most frequently used hashtags in the statically and dynamically obtained Twitter datasets. The dynamic monitor covers a wider range of topics and has a more uniform distribution in terms of the frequency of different keywords. }
\label{fig:compStatic}
\end{figure}

As shown in Figure \ref{fig:compStatic}, the semi-automated dynamic approach is visibly better than a static data collection procedure, producing a more uniform distribution of popular keywords. The dynamic approach keeps old frequently-discussed hashtags (\#inaugurationday) while detecting new viral hashtags (\#trump, \#biden), providing more effective general topic coverage than the static monitor.

\subsection{Observation and Discussion}
We have the following observations in the data analysis and keyword update for this case study.
\paragraph{Word Ambiguity} General terms are ambiguous and may induce irrelevant information. In day 2, we saw a surge in the keyword frequency of ``safety" and ``security" in our filtered corpus. We then decided to include them in our keywords set and expect to capture more text about data security. However, with the knowledge that ``safety" and ``security" are general and ambiguous terms, we take extra caution in the following days. When taking a closer look at the corpus that contains the keywords ``security", we dropped it and include more specific terms like ``datasecurity" and ``databreach" instead. 
\paragraph{Convergence} The dynamic discussion would
``converge" with no new keywords added in the end. In our case study, we did not find new keywords to add to the keyword set in the last three days. Since we did not set a upper limit in the keyword set, it is possible that we captured all the topics after 9 days. In practice, we believe the convergence may or may not happen depending on whether there is a constraint in terms of the number of keywords in total or number of new keywords each round. 
\paragraph{Forecast Information} One difficulty in our case study is the usage of the forecast information. Due to the short time horizon we are working on, there is not enough data for us to train reliable predictive models on the keywords, which impose difficulty in fully automatic update. Fortunately, our method is flexible enough to also account for human intervention. In the future, we aim to further test various type of update rules. 
\section{Related Work}

\paragraph{Information Retrieval}

Existing information retrieval methods like Okapi BM25 \citep{Sparck_Jones_2000} which take a probabilistic approach to ranking documents use a complex and rigid weighting scheme with various parameters that require refinement. Deep learning information retrieval methods like BERT \citep{deshmukh2020irbert} are non-transparent and require abundant data to produce sensible rankings. These approaches focus on optimizing information retrieval in largely static databases and returning results that are most similar or relevant to user's query. Our task is different: social media data is extremely dynamic and prediction intervals occur on smaller scales. Therefore, rankings must be produced efficiently and robustly on relatively small datasets which are streamed real-time. Further, the search process for new keywords begins with a known keyword, but could end in new discoveries which alter the course of data collection. Our target user is uncertain about what they are looking for, and their ``information retrieval'' process is highly informed by real-time predictions and newly exposed keywords. Finally, our application requires flexibility in ranking documents (or keywords): in one use case, corpus frequency may be the strongest indicator of relevance, while in another use case, semantic similarity to an existing set of keywords may be more important.


\paragraph{Word Embedding Models} Previous work has shown that word embedding models serve as an efficient information retrieval mechanism on vast corpora \citep{mci/Galke2017}. In previous work, we show the ability of word embedding models to uncover conversational threads and emergent hashtags in online conversation \citep{liu_etal_2019}. As our application requires real-time processing of streamed social media data, we iteratively train GloVe word embeddings \citep{pennington2014glove} on incoming data to efficiently create a vector representation of the corpus and retrieve information about closest neighbors based on metrics like cosine similarity. To provide greater flexibility and transparency to the ranking scheme, our final implementation sorts keywords based on a linear combination of keyword frequency information and embedding information.

 \paragraph{Keyword Selection}
 
Prior research has focused on the problem of keyword selection \citep{Wang_Chen_Liu_Emery_2016}. This is different from our work in that we assume that the analyst begins with subject-matter expertise that provides keywords and hashtags to seed the dynamic monitor. Other related work has used similar static keyword search query approaches \cite[e.g.][]{oconner:2010,conover_2012,barbera_etal_2015}, which fall short in the study of dynamic debates with rapidly evolving conversation. Another approach uses semi-automated approaches for keyword selection and search by crafting a semi-supervised dynamic keyword methodology \citep{zhengetal2017}. Their approach differs from ours, as we use relatively easy-to-estimate word embedding models and straightforward time-series predictive models, making our approach more intuitive and likely faster computationally. Further, related deep learning methods \cite{Wang_Chen_Liu_Emery_2016} often lack transparency and require vast data to perform well, which precludes their usage in applications with sparse data. Finally, related work has proposed semi-automated keyword selection combining computer and human input \cite{king_lam_roberts_2017}: these are based on complex rules are more effort-intensive, as researchers must themselves apply these rules in their decision process. 
We offer an AI-driven dynamic keyword searching methodology that is fast and efficient (like fully-automated methods), yet provides more transparency and intuition than these methods. We  also offers unprecedented visualizations and a interface for user decision making. 

 
 
 
\section{Conclusion and Discussion}
In this paper, we design and implement a novel dynamic keyword search method for tracking and monitoring fast-evolving online discussions. This closes the gap between the traditional static keyword search method and the highly dynamic data sources in social media. We use word embedding models for finding relevant keyword automatically and use indicators from predictive time series models to help the decision making in keyword updates. We allow for both semi-automatic and fully-automatic data collection. The whole system is build using modern data collection, storage and visualization tools. 

To test the current deployment, we simulate on data collected from the \#MeToo movement and also analyze a practical use case that covers the recent President Inauguration. Our simulation and case study reflects the effectiveness of our framework.

The remaining work on our process will focus on the time series predictive modeling.  While we have deployed our process with a preliminary time series predictive model (which we demonstrate in the \#inauguration case study), we plan to more rigorously test effectiveness of different time series approaches, including traditional linear time series models like ARIMA and deep learning forecasting models like LSTM.


\bibliographystyle{ACM-Reference-Format}
\bibliography{references}

\end{document}